\documentclass[pra,twocolumn,aps]{revtex4}
\usepackage{amsmath}
\usepackage{wrapfig}
\usepackage[pdftex]{graphicx}
\usepackage{wrapfig}
\usepackage[usenames,pdftex]{color}
\usepackage[ansinew]{inputenc}
\usepackage{calrsfs}
\usepackage{amstext}

\definecolor{rem}{rgb}{1.0,0,0}

\begin{document}

\title{Light storage via coherent population oscillation in a thermal cesium vapor}

\author{A.~ J.~ F.~ de Almeida${^1}$, J. ~Sales${^1}$, M.-A.~Maynard$^{2}$, T. Laupr\^{e}tre$^{2}$, F. Bretenaker$^{2}$,  D.~Felinto${^1}$, F. Goldfarb${^2}$, and J.~W.~R.~ Tabosa${^{1,*}}$}

\affiliation{$^1$ Departamento de F\'{\i}sica, Universidade Federal de Pernambuco, 50670-901 Recife, PE - Brazil\\
$^2$Laboratoire Aim\'{e} Cotton, CNRS, Universit\'{e} Paris Sud, ENS Cachan, 91405 Orsay, France\\
 $^*$Corresponding author:~tabosa@df.ufpe.br}


\date{\today}

\begin{abstract}
We report on the storage of light via the phenomenon of Coherent Population Oscillation (CPO) in an atomic cesium vapor at room temperature. In the experiment the optical information of a probe field is stored in the CPO of two ground states of a $\Lambda$ three-level system formed by the Zeeman sublevels of the hyperfine  transition $F=3\rightarrow F^{\prime}=2$ of cesium $D_{2}$ line. We show directly that this CPO based memory is very insensitive to stray magnetic field inhomogeneities and presents a lifetime which is mainly limited only by atomic motion. A theoretical simulation of the measured spectra was also developed and is in very good agreement with the experiment.   

\end{abstract}

\maketitle

\noindent The growing fields of quantum information processing and quantum communications have driven a great interest in the study of coherence preserving memories for light \cite{Lukin03, Lvovsky09}. A common ingredient of any quantum memory lays on the reversible transfer of coherence between light and matter. Different physical phenomena and schemes for implementing such transfer have been demonstrated, including Electromagnetically Induced Transparency (EIT) \cite{Hau01, Lukin01}, Gradient Echo Memories (GEM) \cite{Hosseini09, Buchler12} and  Atomic Frequency Comb (AFC)  \cite{Afzelius09} based memories. The implementation of these memories involves the creation of ground state atomic coherences which are very fragile against atomic motion and magnetic field inhomogeneities. Therefore, an enormous  effort has also been spent to increase the storage time of such memories, where decoherence associated with finite transit time is circumvented by using inert buffer gas or by laser cooling of the atomic sample \cite{Pan09} and decoherence due to inhomogeneous magnetic fields is circumvented by magnetically shielding the sample or by using magnetic insensitive clock transitions \cite{Felinto04, Kuzmich09}. In a recent paper, we have also demonstrated that a five-fold improvement of the storage time of a Zeeman memory in cold atoms could be achieved by applying a dc magnetic field orthogonal to the main magnetic field inhomogeneities \cite{Lezama10}.

On the other hand, it is well known that optical information can also be impressed into the atomic level populations \cite{Cardoso99, Tabosa99}. However, for a closed two level system (TLS) the atomic level population changes, i.e., saturation induced by an optical field, decay with the excited state lifetime which is usually in range of tens of nanoseconds and strongly limits the storage time. Nevertheless, for an open TLS one part of the saturation dynamic, the so called Coherent Population Oscillation \cite{Tan67, Boyd81, Berman88}, decays with the ground state lifetime, which opens up the possibility to store optical information in such systems. Indeed, a spatial optical memory based on CPO was proposed in \cite{Wilson-Gordon10}. Differently from the memories based on ground state coherence, a CPO based memory will be insensitive to magnetic field inhomogeneities, thus eliminating the need of a magnetic field free environment and allowing long storage time  \cite{Goldfarb14}. In this work we first characterize the CPO and EIT spectra by the application of either a constant and uniform magnetic field or a linear gradient magnetic field and demonstrate a new type of optical memory based on CPO at room temperature cesium atoms. 

The experiment is performed in a room temperature cell filled with $10^{-6}$ Torr of Cs atoms, without buffer gas. We employ the degenerate two-level system associated with the cesium hyperfine transition $6S_{1/2}(F=3)\leftrightarrow 6P_{3/2}(F^{\prime}=2)$. The Doppler broadened transition half-width at  half-maximum is about 190 MHz. A simplified experimental scheme is shown in Fig. 1(a). The cesium cell is placed inside three layers of $\mu$-metal shielding and two independent solenoids, one with a constant pitch and the other with a variable pitch, which produce a dc constant and a linear gradient magnetic field, respectively. The $\mu$-metal shielding allows us the systematic control, through the two internal solenoids, of the dc constant and the gradient components of the magnetic field inside the cell. Light from an external cavity diode laser has its frequency first locked to the crossover resonance associated with the transitions $F=3 \rightarrow F^{\prime}=2$ and $F=3 \rightarrow F^{\prime}=4$, as indicated in the hyperfine level scheme shown in Fig. 1(b).  After passing through a $\lambda/2$ wave-plate and a polarizing beam splitter (PBS), the two beams have their frequencies shifted down by two independent Acousto-Optic Modulators (AOM) to nearly resonance with the $F=3 \rightarrow F^{\prime}=2$ transition (852 nm). This allows us to control independently both the amplitude and the frequency of each beam. In order to avoid beam misalignments during the frequency scan, one of the AOMs operates in double passage. The two orthogonal linearly polarized beams, the coupling (C) and the probe (P) beams, are superposed at another PBS and sent into the cesium cell forming a very small angle ($\theta \approx$ 0.2 mrad) in order to minimize the beating due to leakage of the coupling beam. The power of the coupling and probe beams are equal to 120 $\mu$W and 10 $\mu$W respectively before the cell (which is equivalent to averaged Rabi frequencies of 0.4 MHz and 70 kHz respectively inside the cell). Their relative intensity can be controlled by a $\lambda/2$ wave-plate. The two beams have approximately the same diameter of 5 mm. The linear transmission of the low intensity probe is $27\%$. After transmission through the cell, the beams are separated by another PBS and the probe beam is detected by a fast photodiode. 

\begin{figure}[!tbp]
  \centerline{\includegraphics[width=9.9 cm, angle=0]{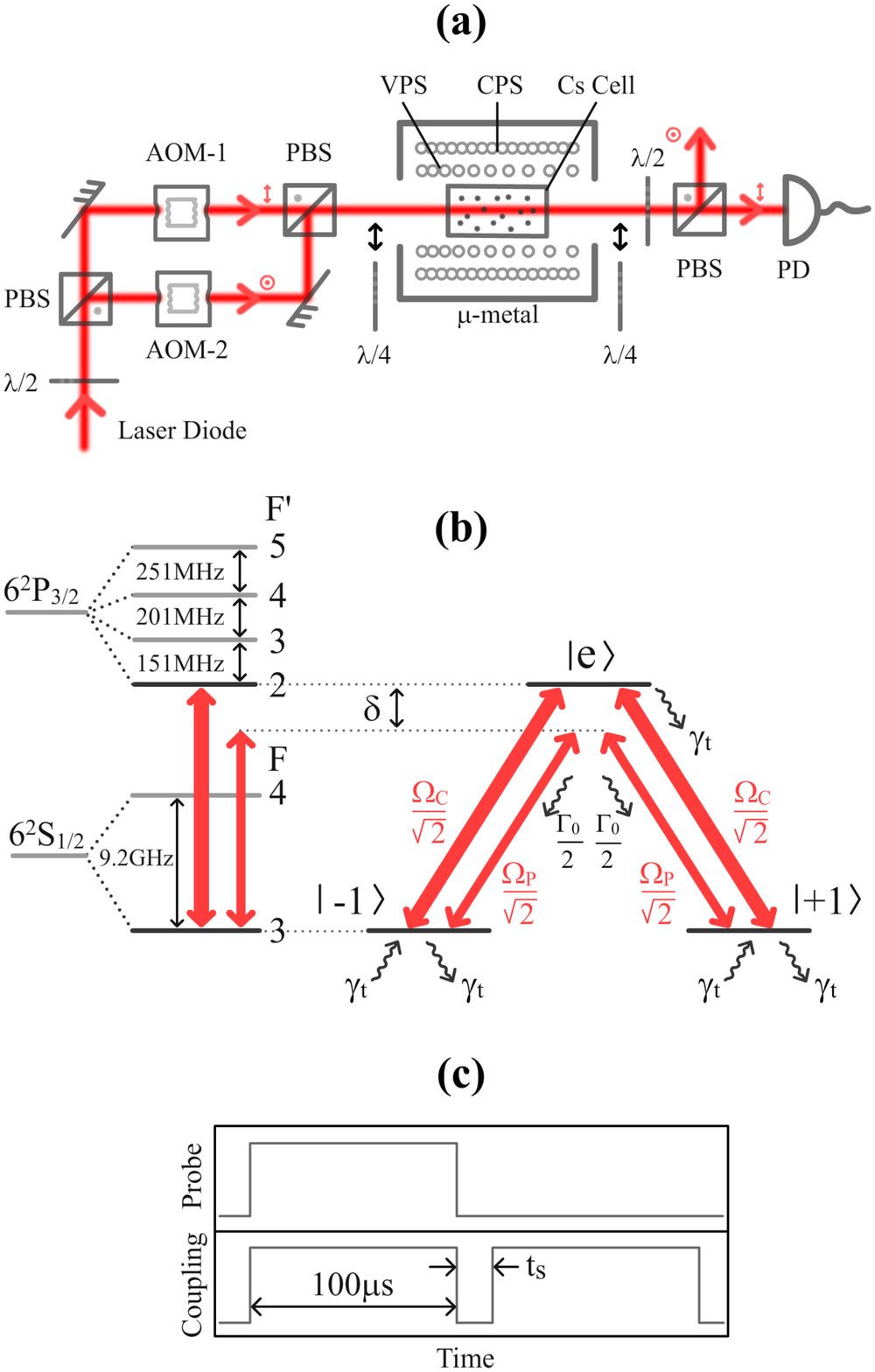}}
  \vspace{-0.5cm}
  \caption{(Color online) (a): Simplified experimental scheme for the observation of CPO/EIT spectra and the associated light storage. PBS (Polarizing Beam Splitter), AOM (Acousto-Optic Modulator), CPS (Constant Pitch Solenoid), VPS (Variable Pitch Solenoid), PD (Photodiode). (b) Hyperfine levels of cesium $D_{2}$ line, showing one generic three-level system associated with Zeeman sublevels of ground and excited states. $\Omega_C$ ($\Omega_P$) is the coupling (probe) Rabi frequency, $\Gamma_0$ is the population decay rate and $\gamma_t$ the transit rate of the atoms into the beams. (c) Time sequence for the writing and reading of the CPO and EIT memories. $t_{s}$ is the storage time.}
  \label{fig:Fig1}
\end{figure}

In Fig. 2 we show the probe beam transmission spectrum when a longitudinal dc magnetic field of magnitude $B=0.9$\,G is applied along the cell. For this spectrum the coupling-probe beam frequency detuning $\delta$ is scanned keeping the  frequency of the coupling beam at the center of the atomic frequency transition for zero magnetic field. 
To understand the observed spectrum we note that for a such orthogonal linear polarization configuration of coupling and probe beams, perpendicular to the applied magnetic field (which defines the quantization z-axis), the two beams have both $\sigma^{+}$ and $\sigma^{-}$  polarization components which excite transitions satisfying the selections rules $\Delta m_{F}=\pm 1$, as indicated in the generic $\Lambda$-three level system shown in Fig. 1(b). As consecutive magnetic Zeeman sublevels of the hyperfine ground state $F=3$ are shifted by $\Delta_{Z}=g_{F}\mu_{B}B$, where $\mu_{B}$ is the Bohr magneton and $g_{F=3}=-1/4$ is the Land\'{e} factor, Raman resonance condition is fulfilled for coupling-probe detuning given by $\pm 2\Delta_{Z}$. Therefore, the two peaks located at $ \approx \pm0.7 MHz$ correspond to the usual EIT resonances. This is consistent with the Zeeman shift predicted for the cesium hyperfine ground state F=3 (equal to -0.35 MHz/Gauss \cite{rabi}) for the applied magnetic field. It is worth noticing that the Zeeman shift in the excited state $F^{\prime}=2$ is equal to -0.95 MHz/Gauss \cite{rabi}, which makes the one-photon coupling detuning slightly different for the other three-level systems associated with different values of the magnetic quantum number $m_{F}$ of the transition $F=3 \rightarrow F^{\prime}=2$. 

\begin{figure}[!tbp]
  \centerline{\includegraphics[width=8cm, angle=0]{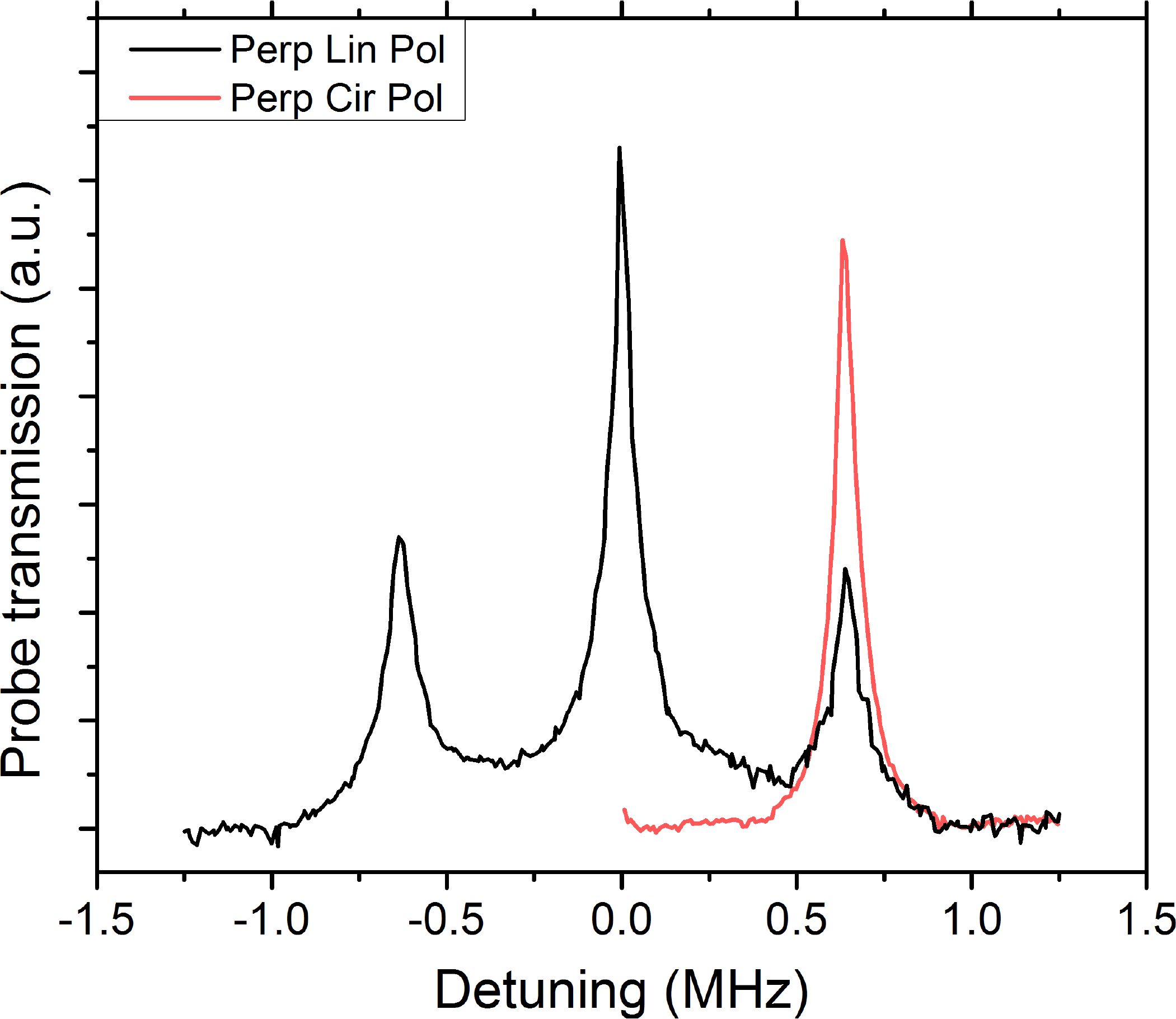}}
  \vspace{-0.2cm}
  \caption{(Color online) Black curve: Measured probe transmission spectrum for orthogonal linear polarization of the coupling and probe beams, when a longitudinal constant magnetic field of magnitude $B=0.9 G$ is applied. The central peak at $\delta=0$ correspond to CPO resonance while the lateral peaks are associated with EIT resonances as explained in the text. Red curve: Measured spectrum under the same experimental conditions, but for orthogonal circular polarizations of coupling and probe beams.}
    \label{fig:Fig2}
\end{figure}

Now we turn our attention to the central peak occurring at $\delta=0$. Clearly it does not correspond to any Raman resonance since the Zeeman degeneracy of the ground state was removed by the magnetic field. In fact this  peak corresponds to CPO resonances, where the same circular polarization components of the coupling and probe beams interact with two different open TLS, with opposite one-photon detunings. 
Indeed, we should note that similar resonances were already observed before in metastable $He$ \cite{Goldfarb12}. To corroborate this interpretation, we have also recorded the probe transmission spectrum when the incident orthogonal linear polarizations are converted into orthogonal circular  polarizations by a $\lambda/4$  wave-plate placed just before the cesium cell. In this case another $\lambda/4$ is also placed after the cell to allow for the separation of probe and coupling beams. The correspondent spectrum is shown by the red curve in Fig.\,\ref{fig:Fig2}, which just presents one single EIT peak located exactly at the same frequency detuning as in the previous spectrum. The difference in the EIT peak amplitudes could be associated with optical pumping induced orientation by the strong circularly polarized coupling beam which essentially selects one three-level system involving more atoms and higher Clebsch-Gordan coefficients, and could also be due to the fact that all the coupling power is now used for this $\Lambda$ system, while it was divided between several $\Lambda$ systems in the previous case. However, the width of the CPO and EIT resonance peaks are of the same order, indicating that both are determined by the same broadening mechanism. 

\begin{figure}[!tbp]
  \centerline{\includegraphics[width= 8.2cm, angle=0]{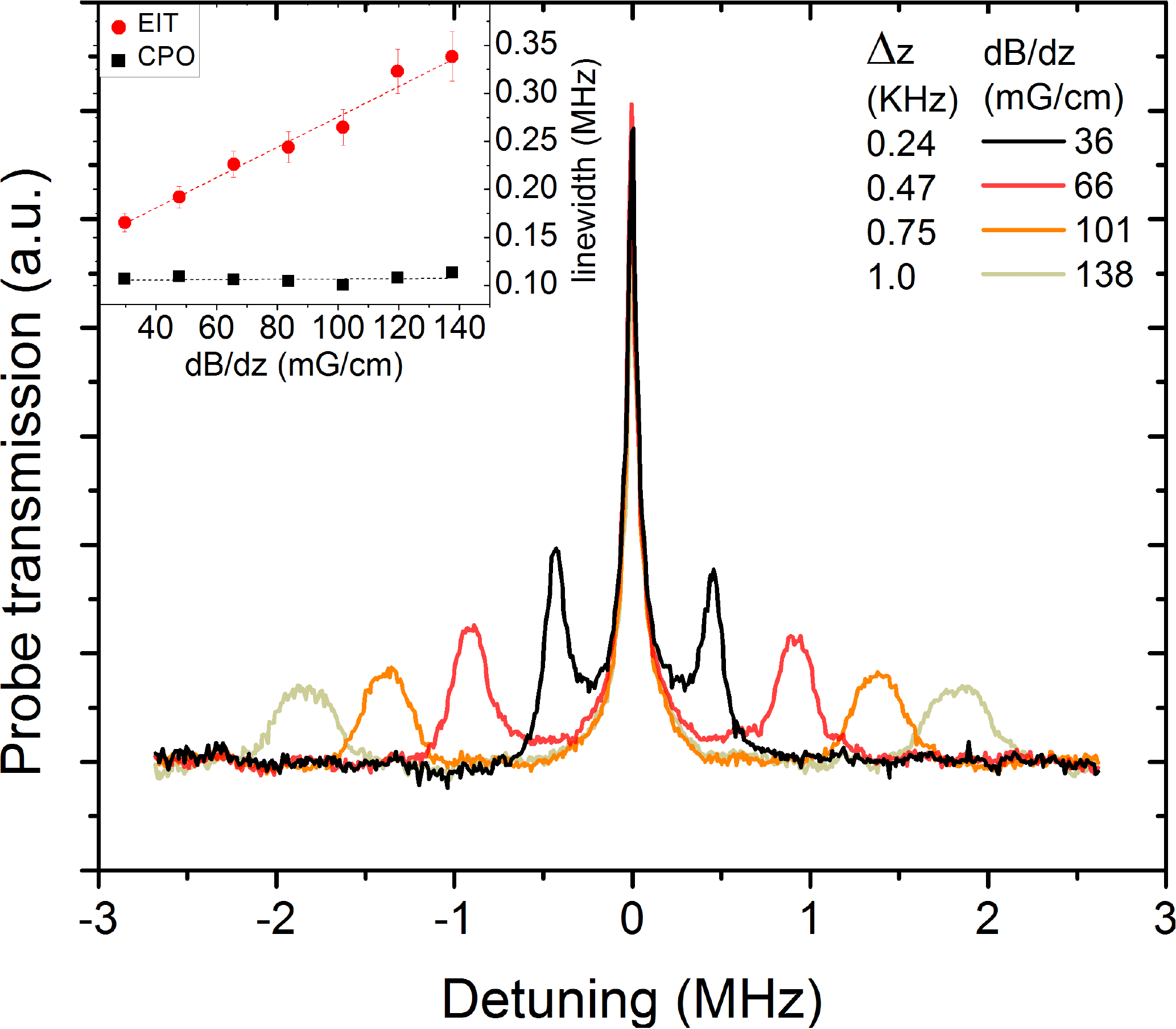}}
  \vspace{-0.2cm}
  \caption{(Color online) Probe transmission spectra for different values of a linear magnetic field gradient (and dc magnetic field) for orthogonal linear polarization of coupling and the probe beams. In the inset we show the linewidth of CPO (black squares) and EIT(red dots) peaks as a function of the linear magnetic field gradient.  }
  \label{fig:Fig3}
\end{figure}

In order to demonstrate that the CPO resonance is not sensitive to magnetic field inhomogeneities, we have also recorded the CPO spectrum for different values of the longitudinal magnetic field gradient. For these measurements, we switched off the constant pitch solenoid and turned on the variable pitch solenoid shown in Fig.\,\ref{fig:Fig1} (a). This last solenoid produces a magnetic gradient (not centered on zero magnetic field) which is mainly linear along the 5 cm cell length. These results are shown in Fig. 3. As the value of the constant magnetic field component is not the same for the different applied magnetic field gradients, the Zeeman shifts $\Delta_{Z}$ and the positions of the EIT peaks change for the different recorded traces. Fig. 3 also clearly shows that the amplitude and width of the CPO peak are not affected by the magnetic field gradient while the EIT resonances are broadened with a decreasing amplitude. In the inset of Fig.\,\ref{fig:Fig3} we plot the measured linewidth for the CPO and EIT resonances as a function of the magnetic field gradient: the linear behavior is consistent with the fact that the two photon resonance occurs for different probe frequencies along the cell.

\begin{figure}[!tbp]
  \includegraphics[width=1.0\linewidth]{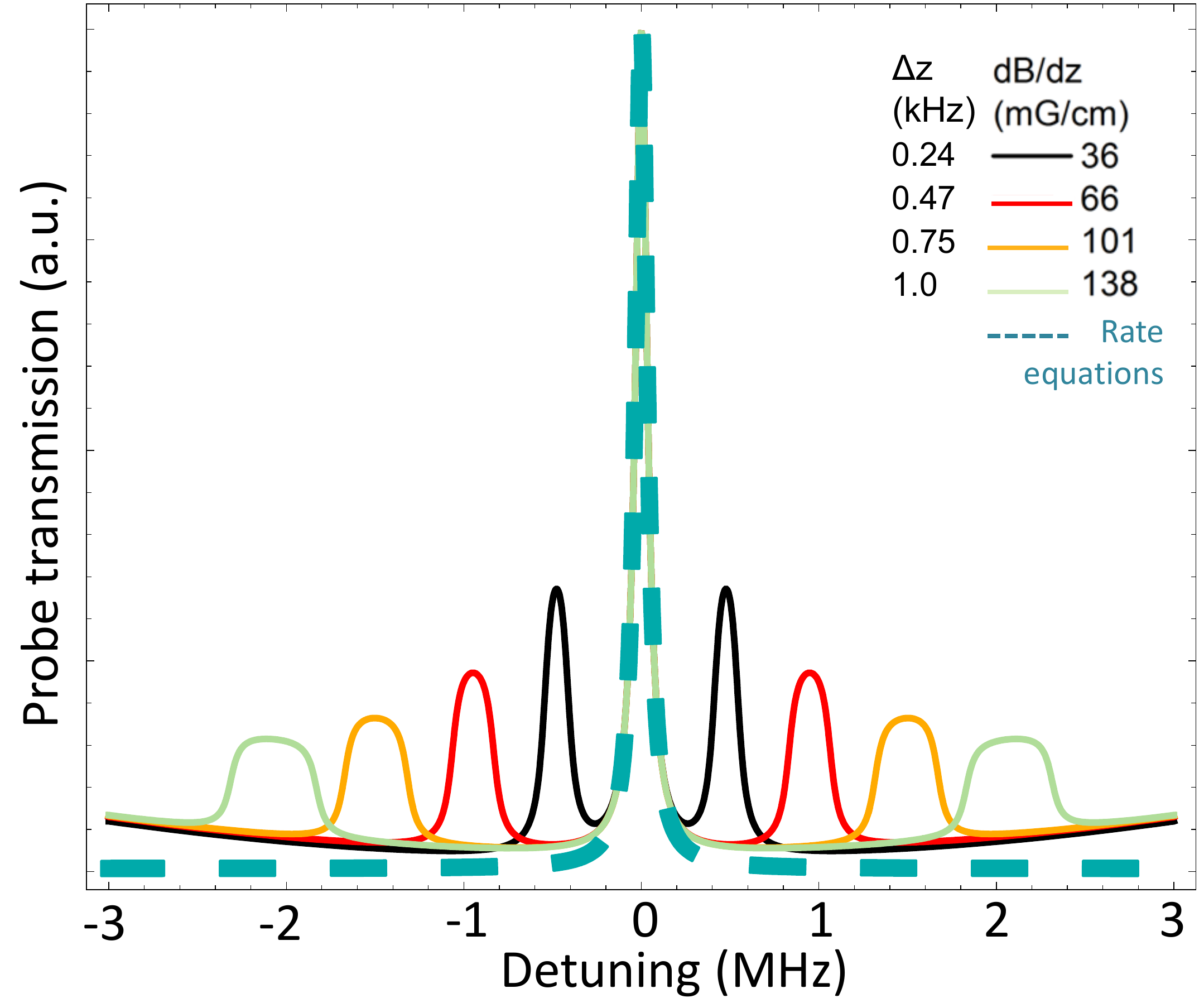}
  \caption{(Color online) Simulated transmission spectra of a three-level system for different values of a linear magnetic field gradient using a Floquet expansion to solve the density matrix equations (continuous lines) or a simple rate equation model (dashed blue line). $\Gamma_0 /2\pi=5.2\,\mathrm{MHz}$, $\gamma_t /2\pi=40\,\mathrm{kHz}$, the optical coherence relaxation rate for one transition $F=3 \rightarrow F^{\prime}=2$ and the Doppler broadening are respectively $\Gamma=\Gamma_0 /2$  and $W_D/2\pi=190\,\mathrm{MHz}$. The averaged coupling and probe Rabi frequencies are $\Omega_C /2\pi=0.4\,\mathrm{MHz}$ and $\Omega_P /2\pi=70\,\mathrm{kHz}$.} \label{fig:Fig6}
\end{figure}

We can try to reproduce these spectra by two different methods, using a density matrix description or rate equations as was done previously in the case of $He^*$ \cite{Goldfarb12}. The first model uses a first-order Floquet expansion of the density matrix of the three-level system. It includes an integration over the magnetic field linear gradient and the results of such calculations are shown as full lines in Fig.\,\ref{fig:Fig6}, for the same values of the magnetic field gradient as in the experiment. The behavior of the EIT resonance shapes, heights and widths are in excellent agreement with the experimental data plotted in Fig.\,\ref{fig:Fig3}. The results obtained by this approach also reproduce the fact that the central resonance is unaffected by the magnetic field inhomogeneities, which indicates again that it is not due to coherences between the ground-state sublevels. Consequently, since it must depend only on the dynamics of the sublevel populations, it should be well taken into account by a rate equation model.

In \cite{Goldfarb12}, the rate equations in the $\Lambda$ system were obtained considering an open TLS, which stands for one leg of the $\Lambda$ system, the other leg being just dealt with as an extra decay channel for the upper level. Here, we derive a more complete rate equation model for the three levels, which describes the two legs of the $\Lambda$-system on an equal footing. If $N_{e}$, $N_{\textrm{-}1}$ and $N_{1}$ are the populations of the upper level $|e\rangle$ and the lower levels $|-1\rangle$ and $|1\rangle$ respectively (see Fig.\,\ref{fig:Fig1} for the notations), one obtains:
\begin{eqnarray}
	\frac{dN_e}{dt} & =&-(\Gamma_0+\gamma_t)N_e+\frac{I^{-}(t)\sigma}{\hbar\omega_0}(N_{\textrm{-}1}-N_e) \nonumber\label{eqn1}\\
&& +\frac{I^+(t)\sigma}{\hbar\omega_0}(N_{1}-N_e) ,\\
	\frac{dN_{\textrm{-}1}}{dt} & =&\gamma_t+\frac{\Gamma_0}{2}N_e-\gamma_tN_{\textrm{-}1}-\frac{I^{-}(t)\sigma}{\hbar\omega_0}(N_{\textrm{-}1}-N_e), \label{eqn2}\\
	\frac{dN_{1}}{dt} & =&\gamma_t+\frac{\Gamma_0}{2}N_e-\gamma_tN_{1}-\frac{I^+(t)\sigma}{\hbar\omega_0}(N_{1}-N_e), \label{eqn3}
\end{eqnarray}
where $\sigma$ is the absorption cross section, $\Gamma_0$ is the decay rate of the upper level and $\gamma_t$  is the decay rate of the lower levels which corresponds here to the transit rate of the atoms through the beam.
$I^+(t)$ is the total intensity (probe plus coupling beams) applied on the transition 
$|1\rangle\rightarrow|e\rangle$ and $I^-(t)$ the total intensity applied on the other leg $|-1\rangle\rightarrow|e\rangle$. With a frequency difference $\delta$ between the probe and the coupling beams, they can be decomposed as:
\[I^\pm(t)=I_0+I^\pm_1e^{-i\delta t}+I^\pm_{-1}e^{i\delta t},\]
with $I^\pm_1\ll I_0$ and $I^\pm_{-1}\ll I_0$. The same first order expansion applies to the populations:
\[N_j(t)=N_{0j}+(N_{1j}e^{-i\delta t}+N_{-1j}e^{i\delta t}),\]
where $j=e,-1,1$. Then the oscillating part of the population inversion $w^\pm(t)=N_e(t)-N_{\pm1}(t)$ is found to be given by:
\begin{eqnarray}
w^{-}_1=-\frac{w_0}{2}\left[\frac{3(I^{-}_1+I^+_1)\frac{\sigma}{\hbar\omega_0}}{\Gamma_0+\gamma_t+3\frac{I_0\sigma}{\hbar\omega_0}-i\delta}+\frac{(I^{-}_1-I^+_1)\frac{\sigma}{\hbar\omega_0}}{\gamma_t+\frac{I_0\sigma}{\hbar\omega_0}-i\delta}\right], \label{eqn4}\\
w^+_1=-\frac{w_0}{2}\left[\frac{3(I^{-}_1+I^+_1)\frac{\sigma}{\hbar\omega_0}}{\Gamma_0+\gamma_t+3\frac{I_0\sigma}{\hbar\omega_0}-i\delta}-\frac{(I^{-}_1-I^+_1)\frac{\sigma}{\hbar\omega_0}}{\gamma_t+\frac{I_0\sigma}{\hbar\omega_0}-i\delta}\right], \label{eqn5}
\end{eqnarray}
where
\begin{equation}
w_0=-\frac{\Gamma_0+\gamma_t}{\Gamma_0+\gamma_t+3\frac{I_0\sigma}{\hbar\omega_0}} \label{eqn6}
\end{equation}
is the dc part of the population inversion. Equations (\ref{eqn4}) and (\ref{eqn5}) exhibit two Lorentzian terms of widths $2(\Gamma_0+\gamma_t)$ and  $2\gamma_t$ at low coupling power. Since the probe and coupling beams are orthogonally polarized, the probe Rabi frequency has opposite signs on the two transitions along the $\Lambda$ system, as shown in Fig.\,\ref{fig:Fig1}. As a consequence, the beatnotes along both legs of the $\Lambda$-system are in antiphase and $I^+_1$ and $I^-_1$ have opposite signs. The first terms in equations (\ref{eqn4}) and (\ref{eqn5}), which correspond to the usual CPO resonance of width  $2(\Gamma_0+\gamma_t)$, are then equal to zero: the CPO resonance width is thus limited by $\gamma_t$, and not by the excited level population decay rate as in usual CPO experiments \cite{Boyd81}.\\
Fig.\,\ref{fig:Fig6} shows in dashed line the resonance that can be obtained with such a rate equation model. It perfectly reproduces the central resonance plotted in Fig.\,\ref{fig:Fig3}, which proves again that it is not a two photon Raman resonance but is in fact a dynamical saturation effect. The CPO storage experiments described below are performed in this transmission window, which does not involve the Raman coherence between the lower levels.

\begin{figure}[tb]
  \centerline{\includegraphics[width= 7.5 cm, angle=0]{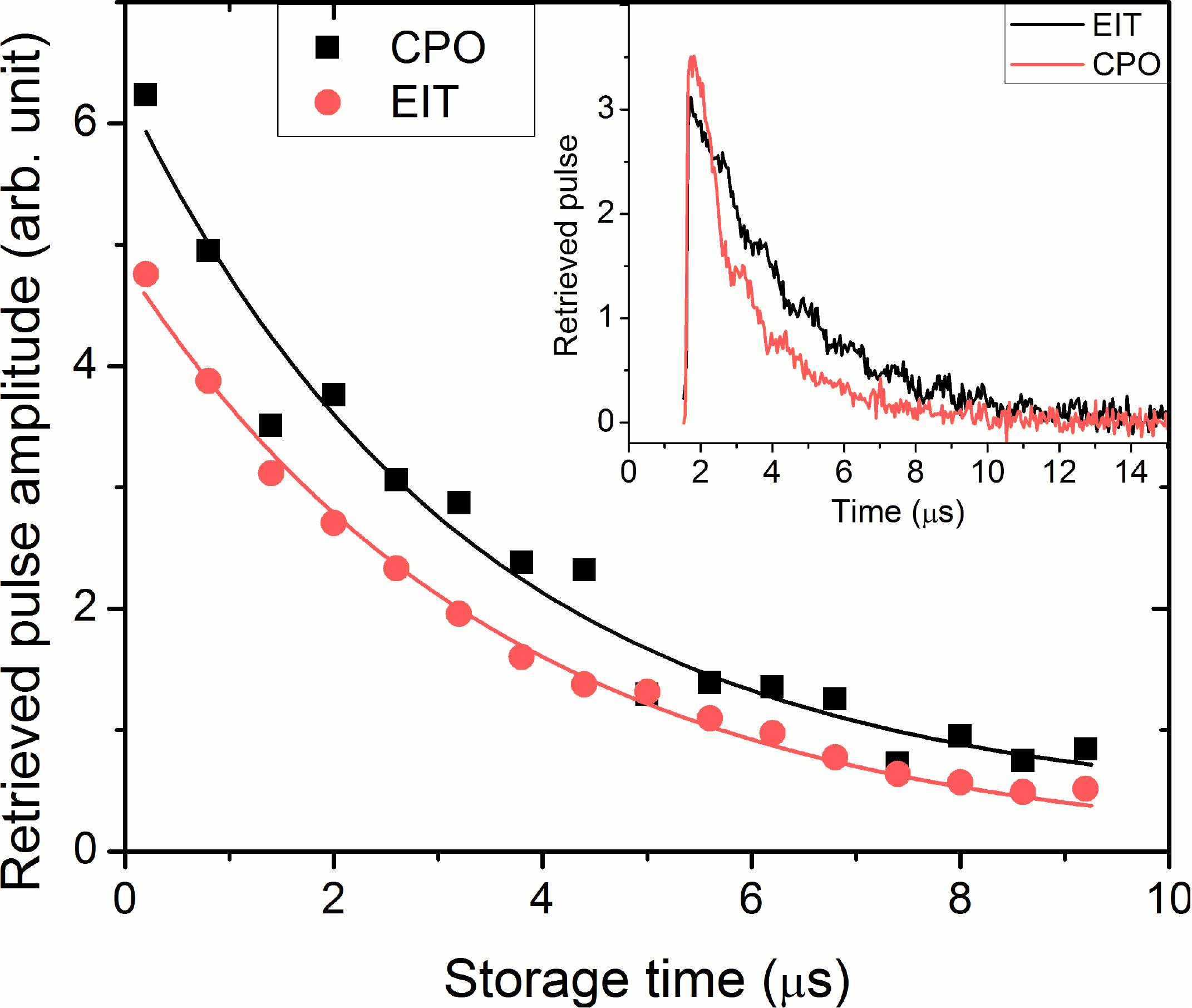}}
  \vspace{-0.2cm}
  \caption{(Color online) Decay of the retrieved pulse amplitude for CPO (black squares) and EIT (red dots) memories in the presence of a constant magnetic field of magnitude B=0.9 G. As explained in the text, the CPO memory uses orthogonal linear polarizations for the coupling and probe beams with $\delta=0$, while the EIT memory employs orthogonal circular polarizations at $\delta=0.7$\,MHz. The solid lines correspond to the respective exponential fittings. In the inset we show the temporal shape of the retrieved pulse, both for the CPO and EIT memories.}
  \label{fig:Fig4}
\end{figure}

In the following we describe the comparative investigation performed on light storage using the CPO and EIT coherent phenomena respectively. For light storage via the CPO resonance, we apply a magnetic field of $B=0.9$\,G and set the coupling-probe detuning at $\delta=0$. We then modulate the amplitude of the coupling and probe beams according to the time sequence shown in Fig.\,\ref{fig:Fig1}(c), where the coupling and probe beams are turned on for about 100 $\mu s$ after which both beams are suddenly switched off. This duration is long enough to allow the CPO to reach a stationary regime. After a controllable storage time $t_{s}$, only the coupling beam is turned back on, and then the retrieved signal pulse is detected with the same polarization as the incident probe beam. As the polarization optics are not perfect, a small leakage of the coupling beam intensity reaches the photodetector, so during the retrieval process one observes the transient oscillations with a frequency determined by the ambient magnetic field  \cite{Lenci12}. In order to eliminate this transient background oscillations, for each storage time, we have recorded the retrieved signal both at the center of the CPO resonance and with the probe beam detuned several linewidths from CPO resonance, where no probe signal is retrieved. We then subtract the two signals to obtain the retrieved pulse as is shown in the inset of Fig.\,\ref{fig:Fig4}. In Fig.\,\ref{fig:Fig4} we show the plot of the retrieved pulse maximum amplitude (black squares) for different storage times. 
Under the same condition of external magnetic field we have also measured the retrieved pulse amplitude for the case of orthogonal circular polarizations for the coupling and probe beams and obtained similar retrieved pulse shapes. In this latter case, the coupling-probe detuning is set to the EIT resonance  peak as shown in Fig.\,\ref{fig:Fig2}. The dependence of the maximum retrieved pulse amplitude with the storage time is also shown in Fig.\,\ref{fig:Fig4} (red dots). Clearly we see that the retrieved pulse amplitudes decay with the same decay time of the order of 3.5 $\mu$s, obtained by exponential fits represented by the solid curves in Fig.\,\ref{fig:Fig4}. The fact the storage decay time is the same for the CPO and EIT memories is consistent with the measurement of the same spectral linewidths of the CPO and EIT resonances: in the absence of magnetic field inhomogeneities, the Raman coherence lifetime is limited by the transit of the atoms through the laser beam, just like the lower level populations. 

\begin{figure}[htb]
  \centerline{\includegraphics[width=7.5 cm, angle=0]{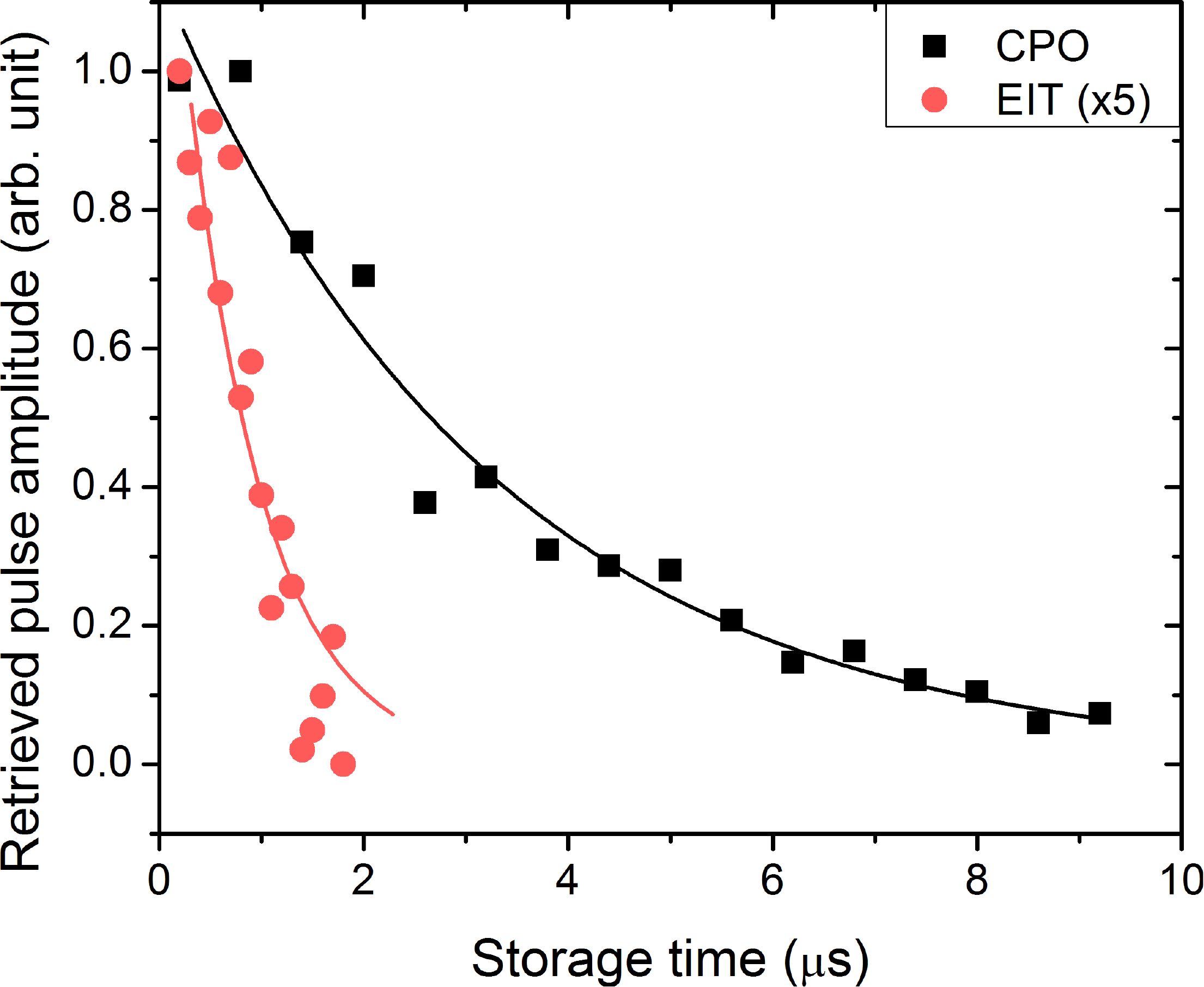}}
  \vspace{-0.2cm}
  \caption{(Color online) Decay of the retrieved pulse amplitude for CPO (black squares) and EIT (red dots) in the presence of a linear magnetic field gradient of magnitude $dB/dz=45$\,mG/cm. The retrieved signal amplitude for the EIT memory is multiplied by a factor of 5.}
  \label{fig:Fig5}
\end{figure}

However, when a linear magnetic field gradient of about $dB/dz=45$\,mG/cm is applied in the cell region, the measured retrieved pulse decay is quite different for CPO and EIT memories, as can be seen in Fig. 6. For the applied linear magnetic field gradient, the EIT resonance is shifted by  approximately the same amount as in the previous case, and in Fig.6 we plot the corresponding retrieved amplitude  for the CPO (black squares) and EIT (red dots) memories. As we can clearly see, now the EIT memory decay time is reduced to $\approx 0.8\,\mu$s, while the decay time of the CPO memory remains unchanged. We have also observed that, contrary to the previous case where the amplitudes of the retrieved pulses are approximately equal, the amplitude of the maximum retrieved signal for the EIT memory is about 5 times smaller than that for the CPO memory in the presence of a magnetic field gradient.   

In conclusion, we have experimentally investigated the CPO and EIT spectra in the presence of a linear magnetic field gradient and developed a theoretical model to describe these spectra either through a Floquet expansion of the density matrix or by a rate equation model, with an excellent agreement between theory and experiment. We have demonstrated a new type of memory for light which is based on CPO in cesium atoms at room temperature. Moreover, an extensive comparative investigation between the CPO and EIT memories under similar environments was performed and the robustness of the CPO memory against magnetic field gradients was demonstrated. We believe this demonstration will be of considerable practical importance since it can be implemented using simpler atomic level configurations, as two-level atoms for instance, and eliminates the necessity of a magnetic free environment.

We acknowledge W. S. Martins for computational assistance in the early stage of the experiment.
This work was supported by the Brazilian agencies CNPq, INCT-QI and FACEPE and by D\'el\'egation
G\'en\'erale de l'armement (DGA), France.

\end{document}